\documentclass[prx, tightenlines,aps,showpacs,nofootinbib,superscriptaddress,longbibliography,twocolumn, 10pt]{revtex4-2}
\usepackage[utf8]{inputenc}
\usepackage{orcidlink}
\usepackage{epsfig}
\usepackage{amssymb}
\usepackage{amsmath}
\usepackage{dsfont}
\usepackage{amsfonts}
\usepackage{dsfont}
\usepackage[toc,page, titletoc]{appendix}
\usepackage{footmisc}
\usepackage{algorithm2e}
\usepackage{bm}
\usepackage{physics}
\usepackage{mathrsfs}
\usepackage{amsthm}
\usepackage{hyperref}
\usepackage[caption=false]{subfig}
\newcommand{\btheta}{{\bm{\theta}}}

\renewcommand{\Re}{\mathfrak{Re}}

\newtheorem{theorem}{Theorem}[section]
\newtheorem{definition}[theorem]{Definition}
\newtheorem{corollary}[theorem]{Corollary}
\newtheorem{proposition}[theorem]{Proposition}

\hypersetup{colorlinks=true, linkcolor=blue, citecolor=blue, urlcolor=blue, unicode=true}
\date{September 24, 2024}
\usepackage{enumitem}
\newlist{abbrv}{itemize}{1}
\setlist[abbrv,1]{label=,labelwidth=1in,align=parleft,itemsep=0.1\baselineskip,leftmargin=!}

\begin{document}
\title{Scalable quantum dynamics compilation via quantum machine learning}

\author{Yuxuan Zhang\orcidlink{0000-0001-5477-8924}$^*$ }
    \affiliation{Department of Physics and Centre for Quantum Information and Quantum Control, University of Toronto,
60 Saint George St., Toronto, Ontario M5S 1A7, Canada}
    \affiliation{Vector Institute, W1140-108 College Street, Schwartz Reisman Innovation Campus
Toronto, \\Ontario
M5G 0C6, Canada}
\author{Roeland Wiersema\orcidlink{0000-0002-0839-4265}$^*$}
    \affiliation{Vector Institute, W1140-108 College Street, Schwartz Reisman Innovation Campus
Toronto, \\Ontario
M5G 0C6, Canada}
    \affiliation{Department of Physics and Astronomy, University of Waterloo, Ontario, N2L 3G1, Canada}
\author{Juan Carrasquilla\orcidlink{0000-0001-7263-3462}$^\dag$}
    \affiliation{Institute for Theoretical Physics, ETH Zürich, 8093, Switzerland}
\author{Lukasz Cincio\orcidlink{0000-0002-6758-4376}$^\dag$}
\affiliation{Theoretical Division, Los Alamos National Laboratory, Los Alamos, NM 87545, United States of America}
\author{Yong Baek Kim$^\dag$}
\affiliation{Department of Physics and Centre for Quantum Information and Quantum Control, University of Toronto,
60 Saint George St., Toronto, Ontario M5S 1A7, Canada}

\begin{abstract}
    Quantum dynamics compilation is an important task for improving quantum simulation efficiency: It aims to synthesize multi-qubit target dynamics into a circuit consisting of as few elementary gates as possible. Compared to deterministic methods such as Trotterization, variational quantum compilation (VQC) methods employ variational optimization to reduce gate costs while maintaining high accuracy. In this work, we explore the potential of a VQC scheme by making use of out-of-distribution generalization results in quantum machine learning (QML): By learning the action of a given many-body dynamics on a small data set of product states, we can obtain a unitary circuit that generalizes to highly entangled states such as the Haar random states. The efficiency in training allows us to use tensor network methods to compress such time-evolved product states by exploiting their low entanglement features. Our approach exceeds state-of-the-art compilation results in both system size and accuracy in one dimension ($1$D). For the first time, we extend VQC to systems on two-dimensional (2D) strips with a quasi-1D treatment, demonstrating a significant resource advantage over standard Trotterization methods, highlighting the method's promise for advancing quantum simulation tasks on near-term quantum processors.
\end{abstract}
\maketitle
\def\thefootnote{$*$}\footnotetext{\noindent Co-first authors}
\def\thefootnote{$\dag$}\footnotetext{\noindent Email address: jcarrasquill@ethz.ch; lcincio@lanl.gov; ybkim@physics.utoronto.ca}
\section{Introduction}
The simulation of quantum dynamics was among the initial applications of quantum computers envisioned by Richard Feynman over four decades ago~\cite{feynman2018simulating}. Today, it remains one of quantum computers' most promising applications, as the exponential size of Hilbert space restricts classical computers. Apart from a pure demonstration of quantum advantage~\cite{preskill2012quantum}, the importance of quantum dynamics simulation is twofold: On the one hand, quantum dynamical processes are of broad interest for physicists in studying problems like the scattering in quantum field theory~\cite{jordan2018bqp,kreshchuk2022quantum}, molecular reactions in quantum chemistry~\cite{kassal2008polynomial,cao2019quantum}, and thermalization and dynamical phase transition in many-body physics~\cite{nandkishore2015many,potirniche2017floquet,yao2017discrete}; on the other hand, the faithful implementation of dynamics is at the core of executing quantum algorithms like 
Quantum approximate optimization algorithm (QAOA)~\cite{farhi2014quantum,farhi2016quantum,guerreschi2019qaoa,zhou2020quantum,zhang2021qed}, quantum phase estimation~\cite{kanno2024tensor}, and may even provide a quantum-enhancement to Markov chain Monte Carlo~\cite{layden2023quantum}.

The conventional approach to simulating quantum dynamics, known as Suzuki-Trotterization~\cite{trotter1959product,suzuki1976generalized}, involves decomposing the unitary evolution generated by a many-body Hamiltonian into small time steps and approximating each step with products of local unitary gates. To achieve a high accuracy in the simulation it is important that each slice is as accurate as possible. While Trotterization is widely used, it is well-known that this type of slicing is far from optimal: For example, randomly reordering Trotter terms can already lead to lower errors~\cite{childs2019faster}. Such inefficiencies raise an important question: can we develop more effective techniques that minimize circuit depth, a critical factor for near-term quantum simulations~\cite{preskill2018quantum}?

Although there are analytical fast-forwarding methods for integrable models that can compress long-time simulations into circuits of constant depth~\cite{cervera2018exact,gulania2021quantum}, it has been proven that these methods do not apply to most generic Hamiltonians~\cite{atia2017fast}. Given this challenge, variational quantum compilation methods present a promising alternative. By leveraging variational optimization, these methods can potentially reduce the gate count and circuit depth while maintaining the accuracy of the quantum simulation, which has been demonstrated by various hardware implementations~\cite{zhang2024observation}. 

\begin{figure}
    \includegraphics[width=0.48\textwidth]{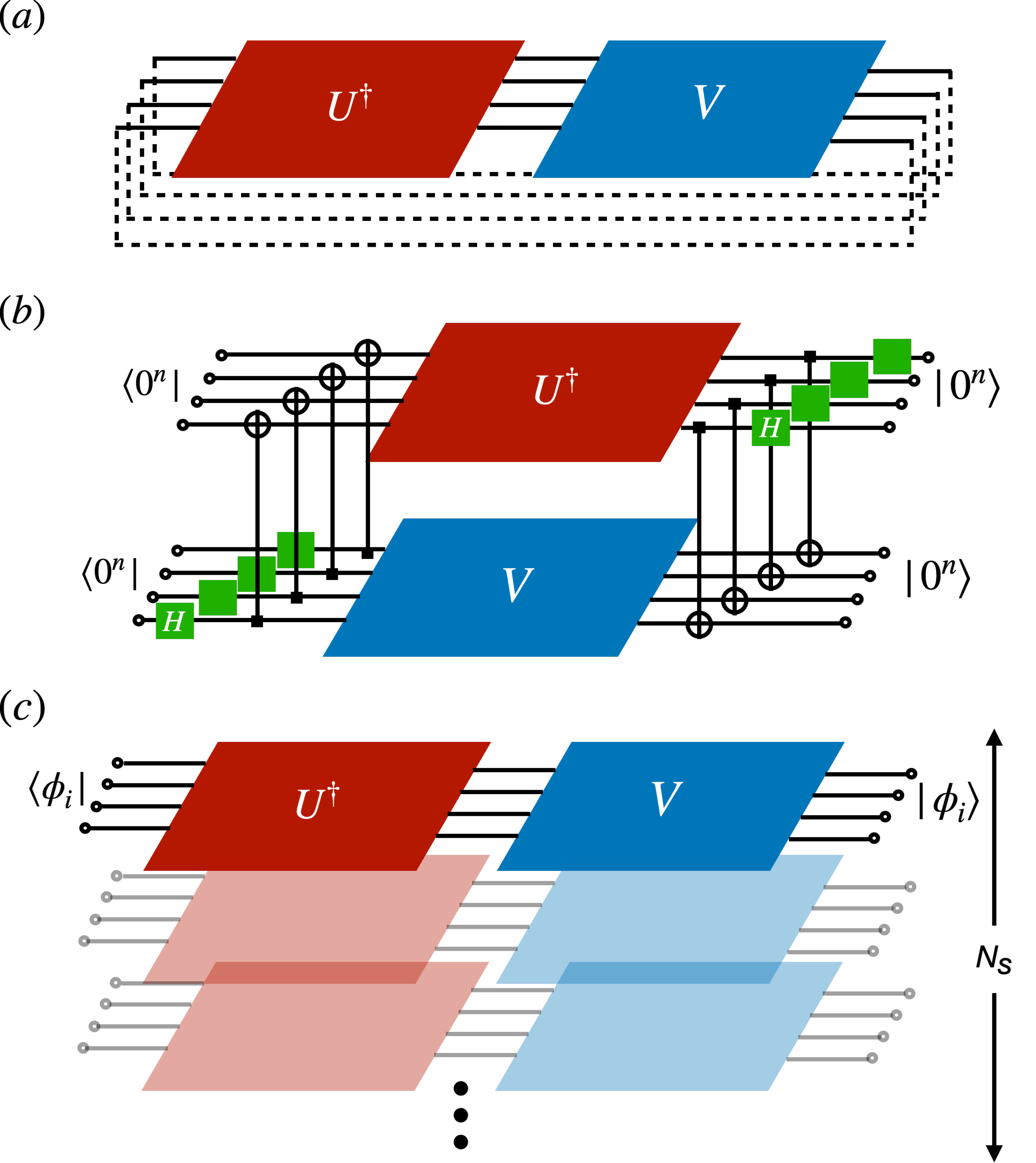}
    \caption{{\bf Tensor-network diagrams for different VQC cost functions.} (a) The most naive cost function for computing two unitaries involves computing $\Tr[U^\dag V]$. This cost function, however, cannot be directly computed on a quantum computer. Additionally, for classical simulations, $U^\dag$ can be costly to write down as a matrix product operator (MPO). (b) The HST circuit is introduced to compute $\Tr[U^\dag V]$ on a quantum computer, and is sometimes considered as a classical cost function~\cite{mizuta2022local}.Green squares represent Hadamard gates. This circuit, however, requires doubling the number of qubits of the systems and introduces highly non-local gates. (c) In this work, we consider a QML-generated cost function, Eq.~\ref{eq:empirical_risk}. The cost function avoids the non-locality introduced by HST and has only a linear dependence on the number of samples $N_s$, which makes it suitable for both NISQ implementations and numerical simulations.In this work, we focus on classical simulation as a primary tool for developing and testing our methods. However, it would be of future interest to extend these techniques to perform compilation on a quantum computer.}
    \label{fig:cost}
\end{figure}

The primary challenge in quantum circuit compilation is accurately evaluating the trace distance between the target unitary operation $U$ and a parameterized quantum circuit (PQC) $V(\theta)$, which is crucial for ensuring that the compiled circuit closely approximates the desired quantum dynamics. First, computing the overlap with conventional methods, such as the Hilbert-Schmidt Test (HST), is both classically and quantumly costly to compute (see Fig.~\ref{fig:cost} for a diagrammatic representation). In fact, for large system sizes, even explicitly writing $U$ as a matrix becomes challenging. Next, variational algorithms that evaluate the full unitary cost face another challenge known as the barren plateau problem, since $U$ and a randomly initialized $V$ are likely to have exponentially vanishing overlaps. As such, early compiling papers~\cite{cincio2018learning,khatri2019quantum,khatri2019quantum,cirstoiu2020variational,lin2021real,yao2021adaptiveVQD} have either been limited by the system size studied, the total time of the dynamics, or focus on the time evolution of a certain given initial state. Additionally, almost all previous works consider $1$D Hamiltonians with little mention of $2$D topologies. 

Some recent works address the challenge by leveraging translation invariance and locality of quantum systems~\cite{mizuta2022local}. However, these approaches still require the use of ancilla qubits, which increase physical errors for hardware implementation of the quantum algorithm and add complexity when running on a classical computer. Others proposed alternative cost functions such as conservation of physical quantities~\cite{zhao2024adaptive}, yet, these works fail to provide any guarantee on the true unitary fidelity, as it is easier to score high on some correlation function benchmark than simulating the dynamics with high fidelity.
\begin{table}[]    
\centering
    \begin{tabular}{c|c|c|c|c}
         $n$& $t$ &  CNOTs & VQC (this work) &Trotter~\cite{barthel2020optimized}  \\\hline
         $4\times 10$ & 0.5 & 550& $\boldsymbol{5.7\times 10^{-4}}$& $4.2\times 10^{-3}$\\
         80& 0.5 & 950 & $\boldsymbol{1.0\times10^{-5}}$ & $1.8\times10^{-2}$ \\

    \end{tabular}

    \caption{For the non-integrable Ising model (Eq.~\ref{eq:ising} at $g = -1$ and $\kappa = 0$) on a 2D rectangular strip and a 1D chain, we compare the generalization risk on random product input states between quantum circuits derived from VQC and those from optimized Trotterization, using the same CNOT gate budget. See Fig.~\ref{fig:cnot} for a detailed comparison.}
    \label{tab:risk}
\end{table}

In this work, we overcome these limitations by leveraging recent advances in quantum machine learning~\cite{Banchi2021generalization, huang2021power,caro2022generalization, caro2023learning}. The key insight is to rephrase the quantum compilation problem as a supervised machine learning task: Suppose, with a PQC, $V$, we want to learn some unitary dynamics $U$, which is provided by a black box that outputs the time-evolved state for any input. The question then arises: what set of input states should we use, and how much data is required to accurately learn $U$? It turns out that the PQC only requires very few input states randomly drawn from the set of all random product states. 
Interestingly, if the trained PQC can learn the dynamics with small errors, then it would not only reproduce the dynamics accurately on samples drawn from the \emph{same} distribution of random product states but also for \emph{other} distributions like the Haar random states: In the machine learning literature, these generalization errors are referred to as the ``in-distribution'' and ``out-of-distribution'' generalization risk~\cite{Valiant1984, bendavid2010learning}, concepts we adopt in this context. Most importantly, the out-of-distribution generalization risk effectively measures the trace distance between $U$ and $V$, which is the ultimate target of VQC.

Previous works~\cite{Banchi2021generalization, huang2021power,caro2022generalization, caro2023learning} have laid the groundwork for employing QML in VQC. However, their numerical simulations were limited to relatively small system sizes. We resolve this issue by employing tensor network techniques: Focusing on short-time quantum dynamics for locally interacting Hamiltonians, we observe that the training process involves computing overlaps between low entanglement states only, which can be computed efficiently with tensor network techniques. Further, inspired by concepts in machine learning~\cite{zhuang2020comprehensive}, we devised initialization strategies that effectively mitigate the barren plateau problem (see Appx.~\ref{app: bp} ), aligning with recent analytical results~\cite{mele2022avoid,drudis2024variational}. The combination of tensor network techniques with a QML-based approach allows us to explore a wide range of system sizes with enhanced performance. Benchmarking with a $1$D Heisenberg model, for the case where $n = 40$ and $t = 0.5$, at the same variational circuit depth ($\tau =10$), our method decreases unitary infidelity from previous VQC results of $\sim0.0023$~\cite{mizuta2022local} to $\sim 0.0009$, which is confirmed with a tensor-network implementation of the HST algorithm. 

For large systems, calculating the infidelity directly between $U$ and $V$ can be computationally hard, especially in higher-dimensions. 
To verify the out-of-distribution generalization of the variationally trained PQCs, we repeatedly applied the circuits to fixed initial states to perform long dynamical simulations. Comparing the time evolution result with dense Trotterization circuits, we find the VQC-generated circuit accurately captures physical correlations. 
Finally, a resource comparison with optimized Trotterization showed that VQC circuits could be orders of magnitude more accurate (see Tab.~\ref{tab:risk}), reinforcing the effectiveness of VQC in the dynamical simulation of large, complex quantum systems across one and higher dimensions.

\section{Results}\label{sec: II}

We consider an $n$-qubit system on a Hilbert space $\mathcal{H} = (\mathbb{C}^2)^{\otimes n}$, which has dimension $N = 2^n$. Quantum computers are expected to excel the most in simulating dynamical problems, where one is interested in simulating the time evolution $U(t) = \exp{-i H t}$ of some state $\ket{\psi}$ generated by some Hamiltonian $H$. Throughout the work, we focus on $1$D nearest neighbor and next nearest neighbor Hamiltonians. 

In all hardware, quantum operations are executed through a set of minimum operations called the universal gate set~\cite{nielsen2001quantum}, usually consisting of single- and two-qubit operations only. Thus, it is a fundamental problem to find a scalable and efficient way of compiling $U$ into products of elementary gates. Conventionally, this could be done deterministically, as canonically exemplified by the Trotterization method~\cite{trotter1959product}. However, analytical and numerical evidence has shown that such deterministic methods are far from optimal. 
In fact, previous results~\cite{atia2017fast,cincio2018learning,khatri2019quantum,berthusen2022quantum} have established evidence that the time evolution of various types of Hamiltonians can be compressible. 

\begin{figure}
    \includegraphics[width=0.45\textwidth]{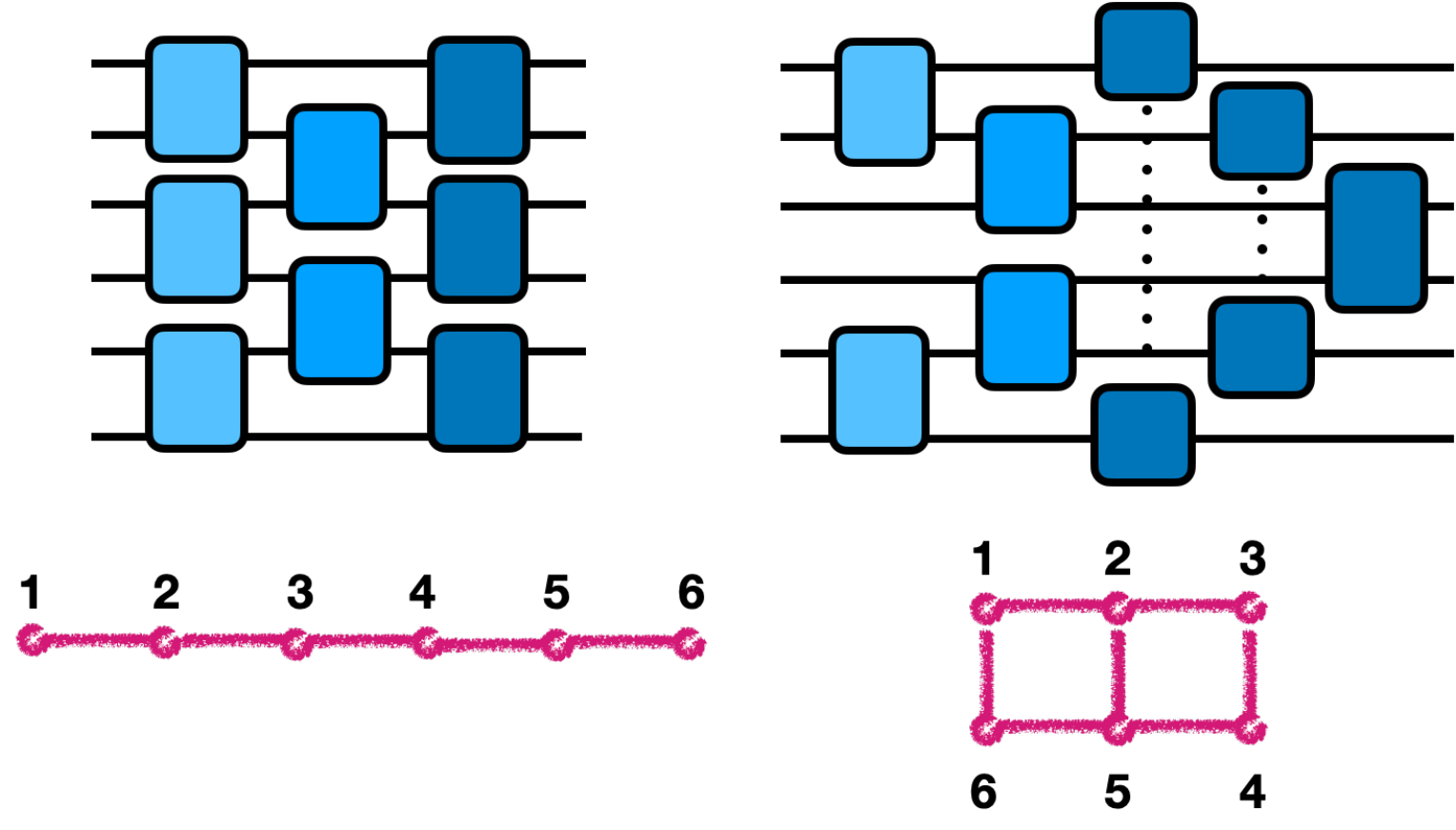}
    \caption{{\bf Geometry of the PQC.} $1$D (left) and $2$D (right) ansatz used for the PQCs are shown here, both with depth $\tau=3$. For $2$D, we use a ``snake" indexing to treat the state as a quasi-1D chain. Each block represents an $\mathrm{SU}(4)$ unitary and unitaries with the same color are considered within the same circuit layer. For the translation-invariant (TI) circuit ans\"atze, we set the parameters in each layer equal to each other.}
    \label{fig:cir}
\end{figure}
\subsection{The QML cost function for learning quantum dynamics}\label{subsec:cost}
For brevity, we now state the main ingredients for our optimization algorithm but will defer the mathematical details to Appx.~\ref{appx:learn}. Given two unitary operators $U$ and $V$, the cost function
\begin{align}
    C(U,V) = 1 - \mathcal{F}(U,V) = 1 - \frac{1}{N^2} \abs{\Tr[U^\dag V]}^2 ,\label{eq:un_fid}
\end{align}
is a measure of distance between $U$ and $V$ where we ignore a global phase. Note that $0\leq C(U,V)\leq 1$ for all $U,V$.

Consider a parameterized unitary circuit $V(\btheta)$ with variational parameters  $\btheta = \{\theta_1,\ldots,\theta_d\}\in\mathbb{R}^d$. We want to approximate $U$ with $V(\btheta)$, for the case where $U = \exp{-i H T}$ and $H$ is some local Hamiltonian. We can achieve this by solving the optimization problem 
\begin{align}
    \min_\btheta\: C(U,V(\btheta)).\label{eq:cuv}
\end{align}
Note that the cost function $C(U,V(\btheta))$ is in many cases prohibitively expensive to evaluate, since for a large number of qubits, the $2^n\times 2^n$ unitary matrices quickly become too large to store in memory. However, recent work in QML theory suggests that one can drastically simplify this optimization procedure~\cite{huang2021power,caro2023learning}. 

First, we can show that the optimization problem of Eq.~\ref{eq:cuv} is equivalent to minimizing the \emph{expected risk}
\begin{align}
    \min_\btheta\: R_{\mathcal{P}_{\mathrm{Haar}}}(\btheta) = \mathbb{E}\left[1 - \abs{\bra{\psi}U^\dag V(\btheta)\ket{\psi}}^2\right]_{\psi\sim \mathcal{\mathcal{P}_{\mathrm{Haar}}}}\label{eq:erisk},
\end{align}
if the test states are drawn from the $n$-qubit Haar random distribution $\mathcal{P}_{\mathrm{Haar}} = \mathcal{S}_{\mathrm{Haar}_n}$~\cite{nielsen2002simple}.
The Haar random states are hard to write down classically, so we simplify the cost function further with three observations:
\begin{enumerate}
    \item Quantum dynamics at a short time has low circuit complexity;
    \item Circuits of low complexity can be learned sample-efficiently with a high generalization accuracy on product states (in-distribution generalization), which are classically easy to compute;
    \item Most importantly, circuits learned with product states will also generalize to Haar random states (out-of-distribution generalization).
\end{enumerate}
The main learning result that establishes the second and the third points comes from ~\cite{huang2021power,caro2023learning}, 
which we restate here:
\begin{proposition}[Equivalence of in- and out-of-distribution risks]
\label{tho:equiv} 
    If a PQC is trained on samples drawn from the random product state, $\mathcal{Q} = \mathcal{S}_{\mathrm{Haar}_1^{\otimes n}}$, then
    \begin{align}
        \frac{1}{2} R_{\mathcal{P}_{\mathrm{Haar}}}(\btheta) \leq \frac{N}{N+1} R_{\mathcal{Q}}(\btheta) \leq R_{\mathcal{P}_{\mathrm{Haar}}}(\btheta).
    \end{align}
\end{proposition}
We call $ R_{\mathcal{Q}}$ the \emph{in-distribution risk} and $R_{\mathcal{P}_{\mathrm{Haar}}}$ the \emph{out-of-distribution risk}. Prop. \ref{tho:equiv} is a powerful result, which states that any PQC trained on the random product states, $\mathcal{Q}=\mathcal{S}_{\mathrm{Haar}_1^{\otimes n}}$ (we use this convention throughout the manuscript), will have an in-distribution risk that is within a factor $1/2$ of the out-of-distribution risk over the Haar measure.
Of course, in a real experiment, one could only obtain a finite number of samples to estimate the expected risk. This empirical estimation is denoted as the \emph{empirical risk}~\cite{caro2023learning}:
\begin{align}
    C_{\mathcal{D}_\mathcal{Q}}(\btheta) = 1 - \frac{1}{N_s} \sum_{i=1}^{N_s} \abs{\bra{\psi_i}U^\dag V(\btheta)\ket{\psi_i}}^2 \label{eq:empirical_risk},
\end{align}
 for a data set $\mathcal{D}_\mathcal{Q} = \{\ket{\psi_i} \}$ of $N_s$ states $\ket{\psi_i} \sim \mathcal{Q}$. As we increase $N_s\rightarrow\infty$, the empirical risk converges to the expected risk, and for finite $N_s$, the empirical risk still yields an upper bound, as we detail in Appx.~\ref{appx:learn}. For a fixed PQC, the tightness of this bound scales as $O(N_s^{-1/2})$, although in practice we find that a small number of states ($N_s<50$) suffices. 

We have reduced the optimization of Eq.~\ref{eq:un_fid} to a QML-based cost function:
\begin{align}
    \min_\btheta C_{\mathcal{D}_\mathcal{Q}}(\btheta)\label{eq:CDQ},
\end{align}
 
\begin{figure*}[htb!]
    \subfloat[Ising, $g = -1, \kappa = 0$, $n=80$, $t=1$]{
    \includegraphics[width=0.37\textwidth]{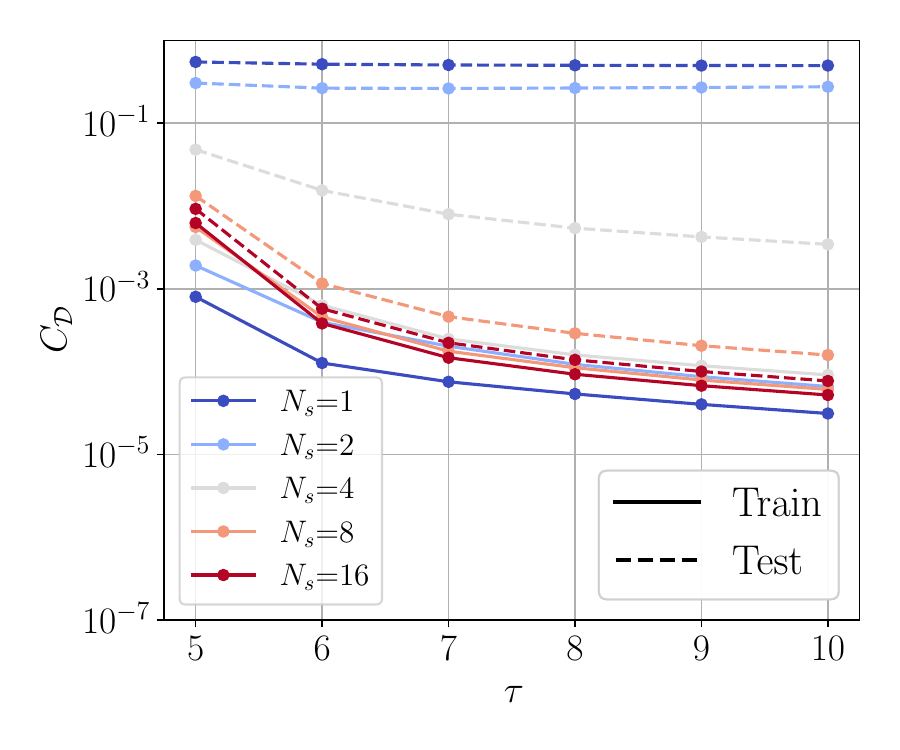}
    }
    \subfloat[Heisenberg, $h = 0$, $n=60$, $t=0.5$]{
    \includegraphics[width=0.31\textwidth]{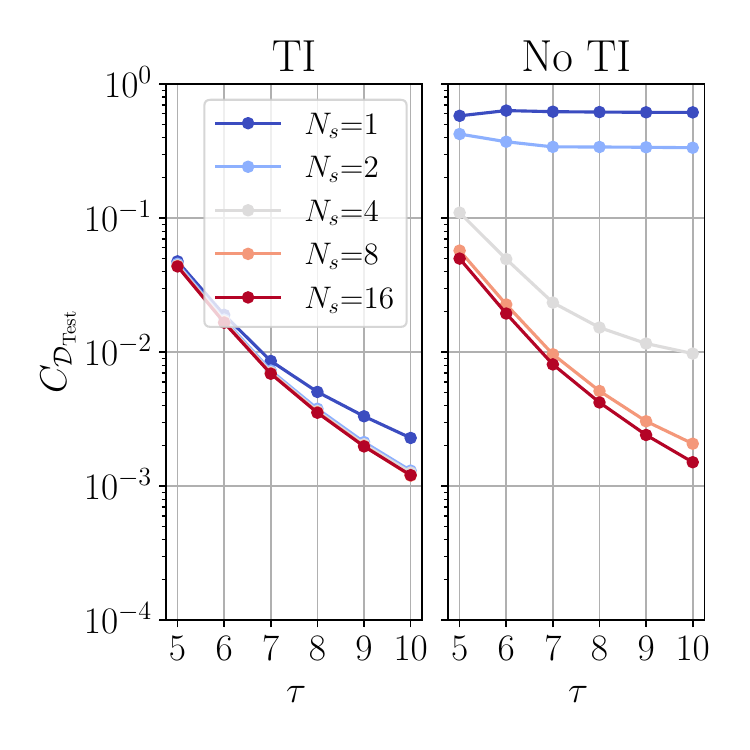}
    }
    \subfloat[Ising, $g = 0, \kappa = 0.2$, $t=1$]{
    \includegraphics[width=0.31\textwidth]{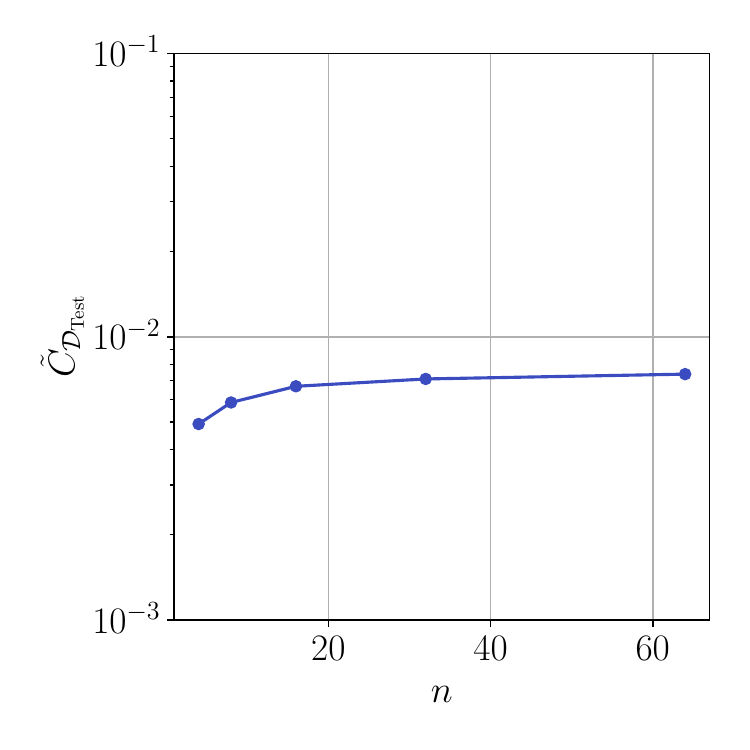}
    }
    \caption{\textbf{In-distribution generalization.} (a) We first train a PQC with no translation invariance (TI) on $N_s$ samples drawn i.i.d. from the random product state ensemble and test it on 100 states drawn i.i.d. from the same distribution. The training and testing loss results are shown on the solid and dashed lines respectively. As we increase the circuit depth $\tau$, the training and testing loss both get lower. As we increase $N_s$, we see that the training loss curves converge, and the testing loss curve approaches the training one. This indicates that an increased number of data samples improves the in-distribution generalization accuracy.
    (b) We compare the testing loss results between PQCs with translation invariance (TI, left) and ones without (right), as defined in Fig.~\ref{fig:cir}. The faster convergence of the training loss curves of the translation invariant PQCs can be reasoned from its smaller number of variational parameters. Additionally, despite the Hamiltonian's open boundary condition, both PQCs converge to similar test losses.
    (c) Fixing the PQC depth at $\tau=4$, we now train our PQC on various system sizes. We observe that, as we increase the system size, the ``per-site'' cost, $\Tilde{C}_{\mathcal{D}_\mathcal{Q}}$, as defined in Eq.~\ref{eq:cost_local} remains relatively unchanged for $n>20$. This shows our VQC results are scalable.}

    \label{fig:indist}
\end{figure*}
Compared with Eq.~\ref{eq:un_fid}, the cost function above, as depicted in Fig.~\ref{fig:cost}c, is significantly more efficient to evaluate on both classical and quantum computers. To see this, first observe that the fidelity function in Eq.~\ref{eq:un_fid} can be computed by either performing a direct contraction or casting the fidelity as an equivalent HST circuit, as illustrated by Fig.~\ref{fig:cost}a and b, respectively. For $1$D and $2$D locally interacting Hamiltonians that we consider in this work, one could utilize tensor-network techniques to compress the time evolution unitary as a matrix product operator (MPO)~\cite{verstraete2004matrix,vidal2004efficient,mcculloch2007density, pirvu2010matrix} (for the contraction in Fig.~\ref{fig:cost}a) or compress a time-evolved state as a matrix product state (MPS) (for the contraction in Fig.~\ref{fig:cost}b and c). However, on the one hand, the bond dimension $\chi$ of an MPO grows faster than the same MPO contracted to a product state; on the other hand, the HST requires not only the same number of ancilla qubits as the system size but also introduces long-range CNOT gates in order to create Bell pairs. The efficiency offered by the QML cost function allows one to probe large system sizes as we compress $U \ket{\psi_i}$ into a high precision MPS with 
time-evolving block decimation (TEBD) method~\cite{verstraete2004matrix,vidal2004efficient}. For many $1$D or quasi-$1$D Hamiltonians, $\chi = \exp(O(t))$~\cite{hastings2007area}, and thus the time evolved states at $t = O(\log(n))$ are asymptotically efficiently computable. 

\subsection{Training and generalization}

Throughout our work, we parameterize the ansatz circuit $V(\btheta)$ with a brickwall structure of general nearest-neighbor $\mathrm{SU}(4)$ gates, as shown in Fig.~\ref{fig:cir}. Since the gradient of the cost $C_{\mathcal{D}_\mathcal{Q}}(\btheta)$ will be exponentially vanishing in the system size $n$~\cite{mcclean2018barren,Cerezo2021costfunctiondep,arrasmith2021effect}, it is crucial that we give the optimization a ``warm start" by initializing the variational gates in a regime of the parameter space where the gradient is barren-plateau-free~\cite{mele2022avoid,ragone2023,drudis2024variational}. Utilizing the structure of quantum dynamics, we outline and compare three different warm start methods to achieve this in Appx. \ref{app: bp}.

To assess the in-distribution generalization properties of the learned circuit $V(\btheta)$, we perform training via gradient descent on a data set $\mathcal{D}_\mathrm{Train}$ with $N_s$ drawn from $\mathcal{S}_{\mathrm{Haar}_1^{\otimes n}}$ and validate afterwards on a data set $\mathcal{D}_\mathrm{Test}$ of $100$ unseen states (also drawn from $\mathcal{S}_{\mathrm{Haar}_1^{\otimes n}}$). This validation set will help us track how well the circuit is performing on in-distribution data.
In Algo. \ref{alg:opt} we show the pseudo-code for our optimization procedure. In practice, we use an early stopping criterion to terminate the optimization if the cost has converged, and we use ADAM~\cite{kingma2015adam} for the gradient descent update step. We provide code  written in quimb~\cite{gray2018quimb} and Pytorch \cite{pytorch} to reproduce our results at 
\cite{our_code}.

\RestyleAlgo{ruled}
\begin{algorithm}[htb!]
\caption{Scalable Variational Compilation}\label{alg:opt}
\KwIn{$N_s$, $H$, $t$, $V(\btheta)$}
\For{$i \in [1,N_s]$}{
    $\ket{\phi_i}\sim \mathcal{S}_{\mathrm{Haar}_1^{\otimes n}}$\\
    $\ket{\psi_i}\gets \mathrm{TEBD}(H,t, \ket{\phi_i})$
}
$\mathrm{step}\gets 0$\\
\While{$\mathrm{step}<1000$}{
    $C_{\mathcal{D}_{\mathrm{Train}}}(\btheta)\gets 1 - \frac{1}{N_s} \sum_{i=1}^{N_s} \abs{\bra{\psi_i} V(\btheta)\ket{\phi_i}}^2 $\\
    $\delta \btheta \gets \nabla C_{\mathcal{D}_\mathcal{Q}}(\btheta)$\\
    $\btheta \gets \btheta - \delta\btheta$\\
    $\mathrm{step}\gets \mathrm{step}+1 $
}
\For{$i \in [1,100]$}{
    $\ket{\phi_i}\sim \mathcal{S}_{\mathrm{Haar}_1^{\otimes n}}$\\
    $\ket{\psi_i}\gets \mathrm{TEBD}(H,t, \ket{\phi_i})$\\
    $C_{\mathcal{D}_{\mathrm{Test}}}(\btheta)\gets 1 - \frac{1}{N_s} \sum_{i=1}^{N_s} \abs{\bra{\psi_i} V(\btheta)\ket{\phi_i}}^2 $
}

\end{algorithm}
 We examine the accuracy of training for two widely-studied physical models: A Heisenberg model with on-site disorder~\cite{vznidarivc2008many,bardarson2012unbounded,serbyn2013universal}

\begin{align}\label{eq:heis}
    H_{\mathrm{Heisenberg}} &= \sum_{\langle i,j\rangle}\left(  \sigma_i^x \sigma_{j}^x +\sigma_i^y \sigma_{j}^y +
    \sigma_i^z \sigma_{j}^z\right)+ \sum_{i=1}^{n} h_i \sigma_i^z, 
\end{align}

where $h_i \sim i.i.d. [-h, h]$; as well as the transversal field Ising model with uniform on-site and next-nearest neighbor perturbations:
\begin{align}\label{eq:ising}
    H_{\mathrm{Ising}} &= - \left( \sum_{\langle i,j\rangle}\sigma^x_i \sigma^x_{j} + \sum_{i=1}^{n}\sigma^z_i \right) + g\sum_{i=1}^{n}  \sigma_i^x 
    \notag\\ &+ \kappa \left(\sum_{\langle i,j\rangle}\sigma^z_i \sigma^z_{j} + \sum_{\langle\langle i,j\rangle\rangle} \sigma^x_{i} \sigma^x_{j}\right).
\end{align}

\begin{figure*}[htb!]
    \subfloat[Heisenberg, $h = 0$]{
    \includegraphics[width=0.33\textwidth]{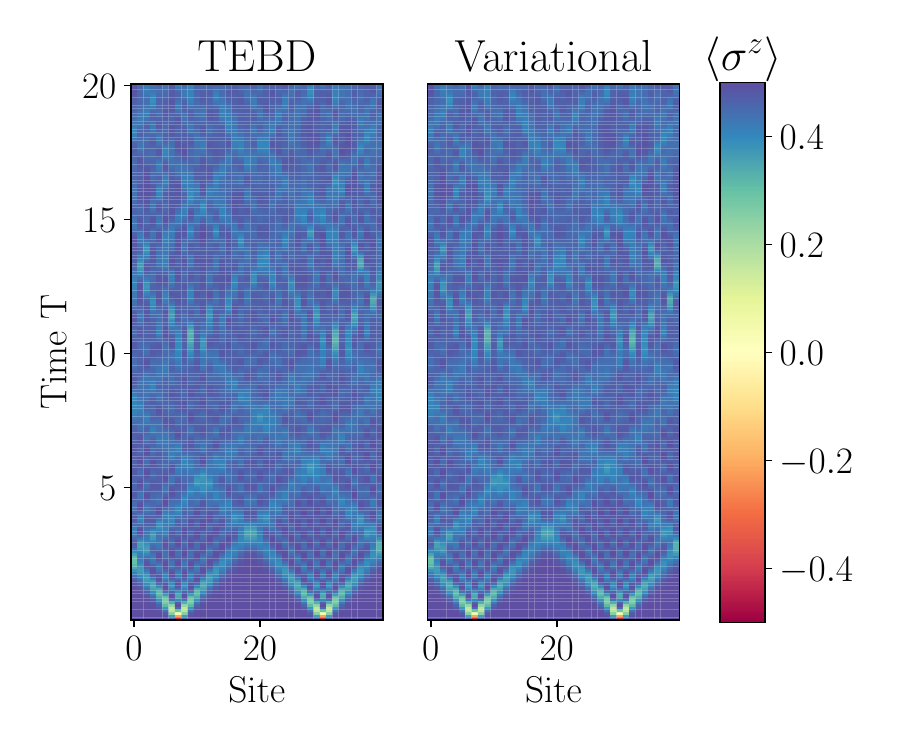}
    }
    \subfloat[Heisenberg, $h = 1$]{
    \includegraphics[width=0.33\textwidth]{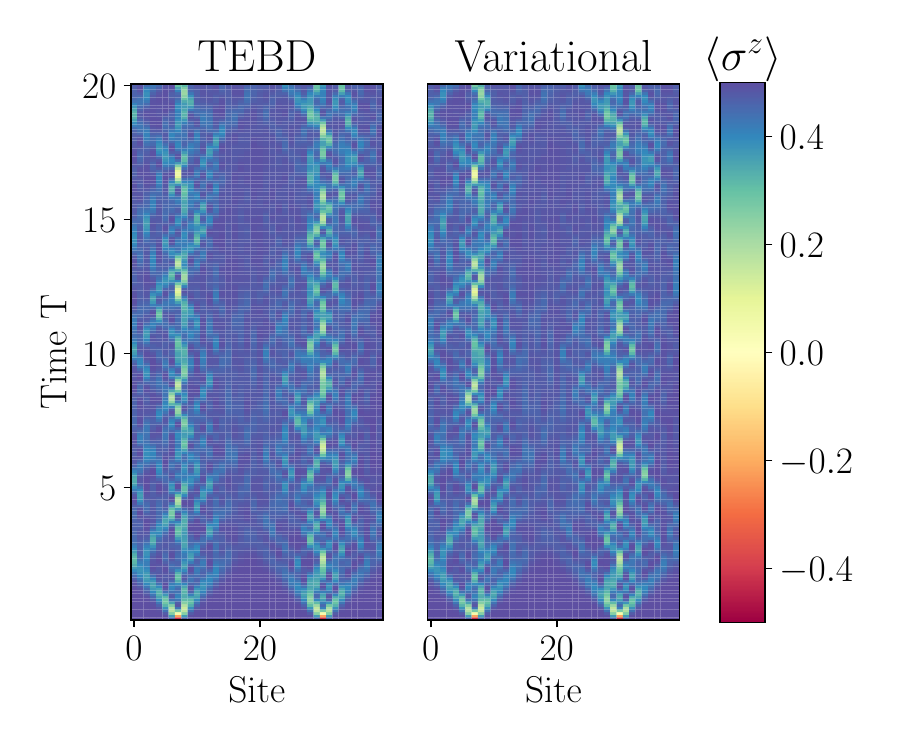}
    }
    \subfloat[Heisenberg, $h = 0$]{
    \includegraphics[width=0.33\textwidth]{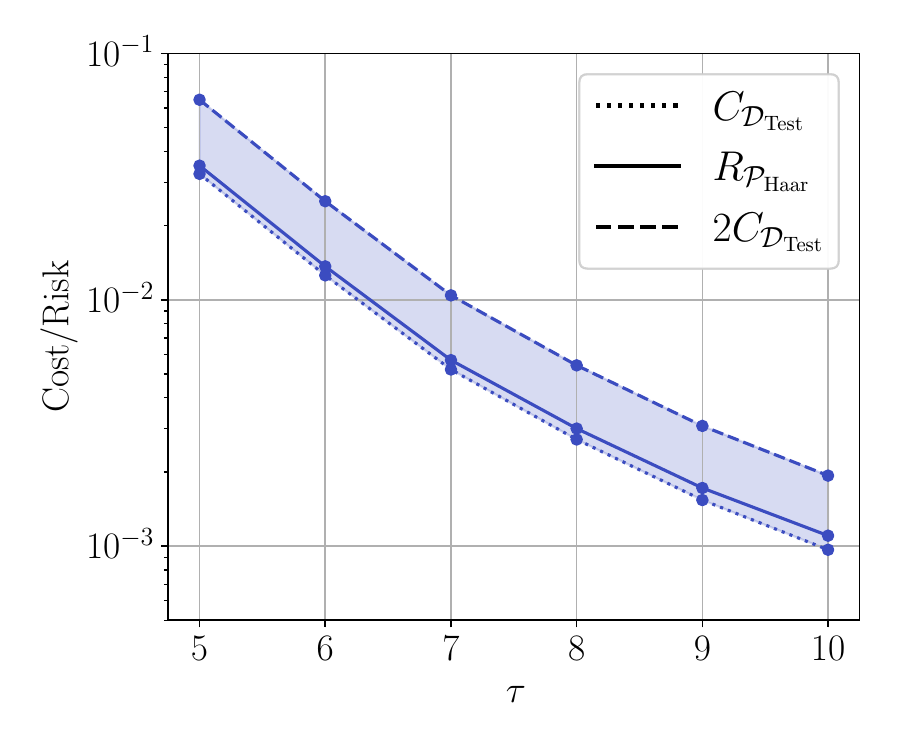}
    }
    \caption{\textbf{Out-of-distribution generalization.} To test out out-of-distribution generalization, we repeatedly apply a VQC trained at time $t=0.1$ to perform a long-time evolution of a computational basis state for total evolution time $T = 20$. We see that we can obtain accurate dynamics at long times for an $n=40$ (a) Heisenberg chain and (b) Heisenberg chain with random fields on each spin. The fidelity of the variational simulation drops from about $\approx 0.99999$ to $\approx 0.99$ between $T=0.1$ and $T=20$, respectively. (c) Using the HST, we compare the actual unitary infidelity to the upper and lower bounds of Prop. \ref{tho:equiv} for the $n=40$ Heisenberg model with $t=0.5$. We see that the test cost is close to the true Haar random risk in practice.} 
    \label{fig:outdist}
\end{figure*}

\begin{figure}
    \subfloat[Magnetization]{
    \includegraphics[width=0.25\textwidth]{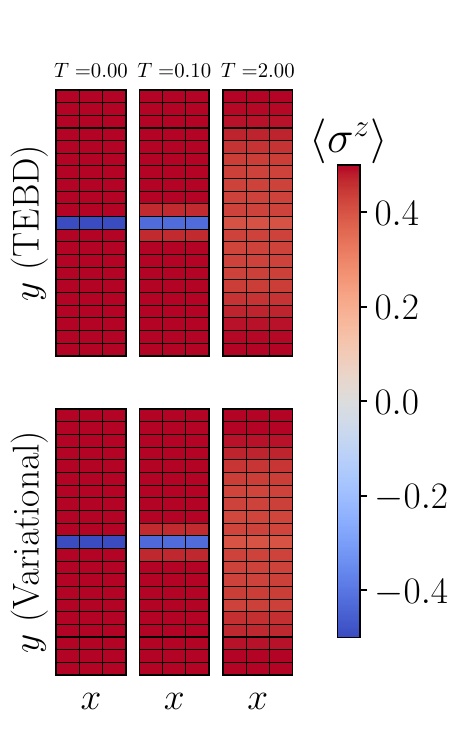}
    }
    \subfloat[Structure Factor]{
    \includegraphics[width=0.25\textwidth]{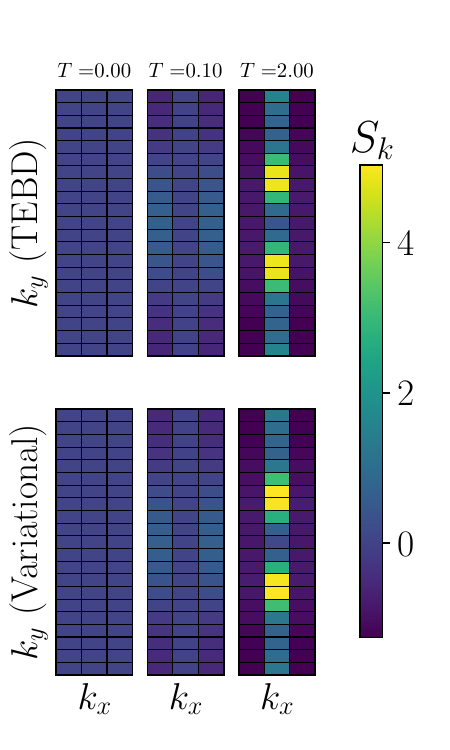}
    }
    \caption{\textbf{Dynamics of $2$D Heisenberg model on a cylinder with a variationally compiled circuit.} Here we target the simulation of the expansion dynamics of ultracold hardcore bosons on a quasi-2D optical lattice after suddenly turning off a strong confining potential~\cite{jreissaty2013, hen2010strongly}. We use the mapping between the hard core boson and spin operators, $S^+ = b^{\dagger}, S^- = b, S^z = b^{\dagger} b - 1/2$. We compile the time evolution of a $t=0.1$ Heisenberg model on a $3\times 21$ lattice with periodic boundaries in the $x$ direction. Then, we quench the state $\ket{0\ldots111\ldots0}$, where the $\ket{1}$ states are exactly in the middle of the lattice corresponding to the confined bosonic cloud. (Left) In real space, we see that the hardcore bosons rapidly expand away from the center of the lattice on both directions. 
    (Right) We calculate the structure factor $S_k = \sum_{\boldsymbol{r}} e^{i \boldsymbol{r}\cdot \boldsymbol{k}} \expval{S^+_j S^-_j}$ for all time steps and show that the dynamics of the momentum distribution exhibits two peaks corresponding to the expanding wings moving up and down in the lattice at roughly constant velocity.}
    \label{fig:2ddyn}
\end{figure}
We start by examining the in-distribution generalization for $1$D systems with open boundary conditions.
Three trends emerge when examining the three panels shown in Fig.~\ref{fig:indist}: 
First, fixing the ansatz to be non-translation invariant, the difference between the objective function and the in-distribution risk decreases as one increases the number of samples. Moreover, perhaps surprisingly, panel b shows that, despite the open boundary condition of the $1$D chains, the translation invariant PQC's generalization outperforms that of the non-translation invariant one due to its fewer parameters. It appears that the extra parameters of the non-TI PQC did not increase its accuracy.
Lastly, fixing the total evolution time and PQC depth while varying the system size, we plot a ``per-site" risk: 
\begin{align}
    \Tilde{C}_{\mathcal{D}_\mathcal{Q}} = 1- (1-C_{\mathcal{D}_\mathcal{Q}})^{1/n},
    \label{eq:cost_local}
\end{align}
against $n$. We find that the per-site risk function flattens already at $n>20$, showing that the QML method can be extended to larger systems even in the presence of next-neighbor interaction. 

Prop.~\ref{tho:equiv} gives a theoretical lower bound for out-of-distribution generalization: for any Haar random ensemble, the risk of generalization (which is equivalent to the unitary cost $C(U,V)$) is at most twice the in-distribution risk. One could verify this by numerically computing $C(U,V)$ for smaller systems, but this quickly becomes infeasible as explained in Sec.~\ref{subsec:cost}. Instead, here we test out-of-distribution generalization by using our compiled unitary to simulate the long-time dynamics of the $1$D and $2$D systems. We also assess the quality of the final compilation using an HST between the model $V(\btheta)$ and a high-accuracy Trotterization of the target $U$.

In Fig.~\ref{fig:outdist} we compare the dynamics generated by the Heisenberg model with and without disorder, starting with a product state $\ket{0...10...10...}$, i.e., a state analogous to two localized particles (the two spins up) in a vacuum. Comparing the magnetization plots between a TEBD simulation and the VQC circuit, we find that the VQC result accurately captures the localization of the particles (spins up) induced by disorder. In panel (c), we also illustrate the bound of Eq.~\ref{tho:equiv} and show that the true infidelity $C(U,V)$ is close to the test risk $C_{\mathcal{D}_\mathcal{Q}}$ in practice, making it a suitable cost function for VQC.

Similarly, we compile the Heisenberg model on a cylinder where the initial state is a stripe of spins down in a sea of spins up followed by time evolution under the Heisenberg Hamiltonian (see Fig.~\ref{fig:2ddyn}). This simulates a system of strongly interacting bosons expanding on an optical lattice, similar to the theoretical and experimental studies in Ref.~\cite{PhysRevLett.94.240403,hen2010strongly,PhysRevA.84.043610,jreissaty2013,wilsonObservationDynamicalFermionization2020}.
We demonstrate the quench dynamics of such a system in Fig.~\ref{fig:2ddyn}.

\subsection{Resource comparison}
One immediate application for VQC is to reduce the number of elementary gates for dynamical simulations compared to Trotterization. Fixing the interaction graph, a variety of methods exist to Trotterize a Hamiltonian \cite{Childs2021theorytrotter, Paeckel2019trotter}. Nevertheless, for Hamiltonians with a particular graph structure and a small number of non-commuting terms, finding the optimal Trotterization at a certain $p$ is possible. For the $1$D nearest neighbor geometry we consider in this paper, we used the optimized parameters in~\cite{barthel2020optimized} for two and three non-commuting Hamiltonian terms, which we use for a fair comparison to our variational compilation scheme. We compare the estimated gate cost between VQC and Trotterization circuits by implementing Ising Hamiltonian dynamics with various parameters and geometries. We report the gate cost in the number of nearest-neighbor CNOT gates in Tab.~\ref{tab:cnot_costs}. 

\begin{figure}[htb!]
    \includegraphics[width=0.4\textwidth]{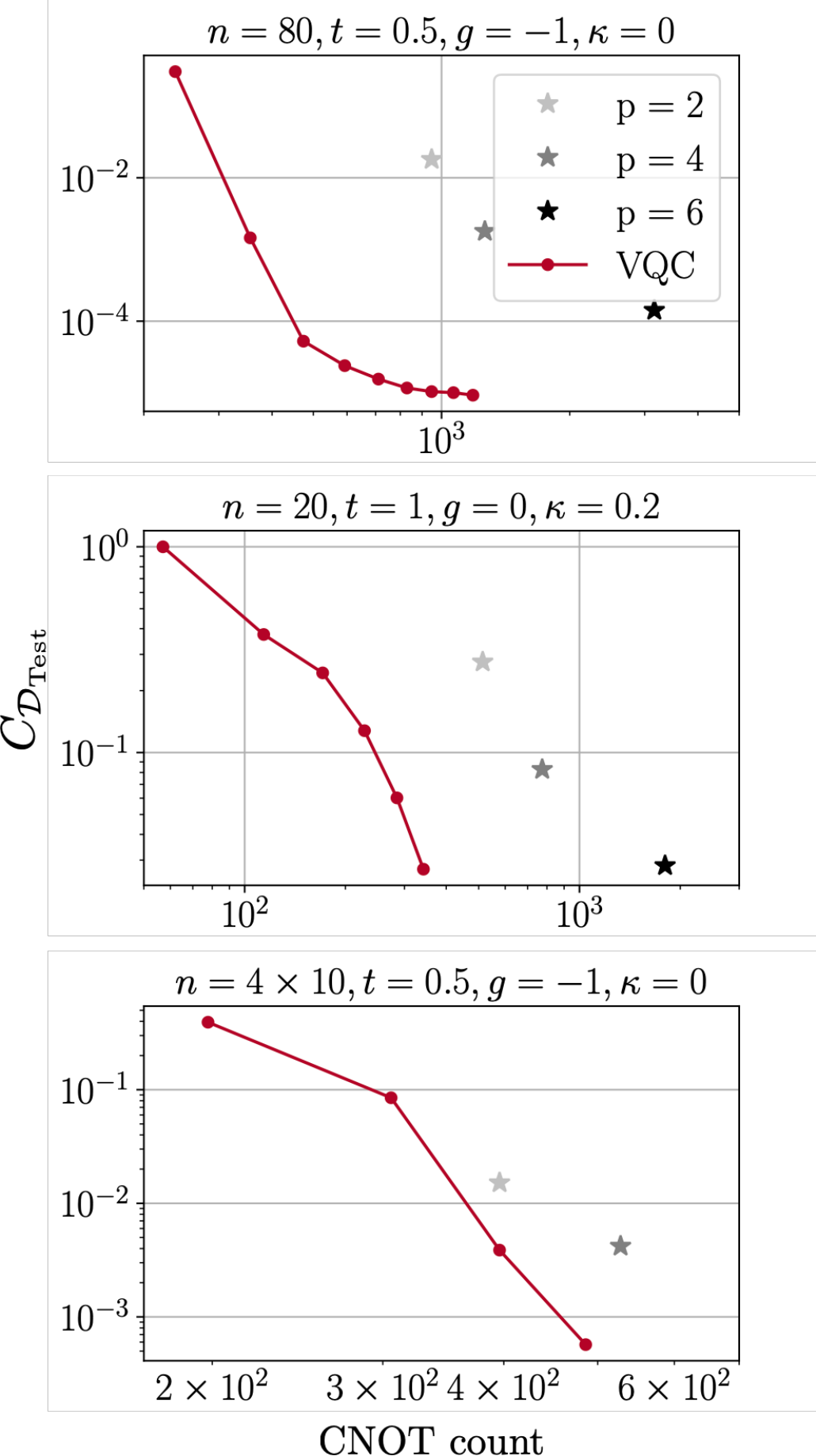}
    \caption{\textbf{A resource comparison.} Here we show a resource and accuracy comparison between implementing quantum dynamics with VQC and optimized Trotterization~\cite{barthel2020optimized} of various orders $p$~\footnote{A Trotterization of order $p$ is such that the leading term of the error grows with $\sim t^{p+1}$.}, tested in $1$D (top and middle) and $2$D (bottom) Ising models. Compared to the Trotterization circuits at the same nearest-neighbor CNOT cost, the VQC circuit would generate circuit with significantly lower generalization risk on random product states, and thus have lower unitary infidelity.}
    \label{fig:cnot}
\end{figure}

Fig.~\ref{fig:cnot} shows that for the same test risk, the QML-compiled circuit requires significantly fewer CNOT gates as compared to Trotterization. In the $1$D case with nearest neighbor interactions only, the PQC's generalization risk already drops below $10^{-4}$ with merely 500 CNOTs, whereas the $p=6$ Trotterization reaches similar risk with a gate count of over 3000. Turning on next-nearest neighbor interactions and setting $t = 1$, the variational circuit matches the $p=6$ Trotterization result with $1/5$th of the gate cost. In $2$D, the resource cost difference between VQC and Trotterization shrinks, which could be the result of the faster scrambling in $2$D geometries or due to the finite-size effect of our system. Nevertheless, compared with $p=4$ Trotterization, the PQC-learned circuit still reduces the generalization risk by a factor of 7 while using fewer CNOT gates.

\section{Discussion}
In this work, we develop a scalable quantum dynamics compilation algorithm for both 1D and quasi-1D systems by reformulating the compilation task as a machine learning problem, employing tensor-network techniques, and introducing novel initialization strategies. Extensive simulations demonstrate that the learned unitary circuits can accurately reproduce long-time dynamics and significantly reduce computational costs compared to highly optimized Trotterization methods. Our integration of quantum compilation with novel simulation techniques holds great potential for near-term quantum computing applications.

Our work also opens up several promising directions for future research. The first natural question is how to extend our method to larger system sizes, particularly in 2D and above. One possibility is to adopt a local update scheme, where each layer of the quantum circuit is optimized sequentially, similar to the approach used in density matrix renormalization group algorithms~\cite{verstraete2004renormalization} for solving ground states. However, as we demonstrate in Appx.~\ref{appx:update}, local updates can sometimes struggle to find the global minimum of the objective function. Nevertheless, the evaluation of the cost function could be improved by optimizing the contraction order of the tensor network. For instance, recent advances in tensor network contraction via belief propagation offer a potential alternative for compiling quantum dynamics on tree-like topologies~\cite{Tindall2023}. On the other hand, instead of the quasi-1D treatment, one may consider representing the time-evolved state with methods that are more natural to higher physical dimensions such as the 2D isometric tensor networks~\cite{zaletel2020isometric}. Our algorithm can directly benefit from the incorporation of these tensor network techniques. 

Next, symmetries naturally emerge in physical systems, as illustrated by the $U(1)$-conserving Heisenberg model considered in this work. An important question is whether these symmetries can be exploited to further enhance compilation quality. Recent advancements in variational quantum computing have explored incorporating symmetries when constructing circuit ansätze~\cite{Meyer2023sym, Nguyen2024equivariant, Wiersema2024geometric}, but can such symmetries be leveraged in the context of compilation? Intuitively, learning the dynamics within a specific charge sector seems more straightforward than learning the entire unitary, as less information is to be captured. This raises the next important question: what is the appropriate ensemble to use for training? To our initial surprise, as proven in Appx.~\ref{app:u1}, we show that training samples for Heisenberg dynamics cannot be generated by any $U(1)$-conserving random circuit with depth less than $n$, as such circuits fail to form a local scrambling ensemble, even within the desired charge sector. This contrasts sharply with the non-charge-conserving case, where a depth-1 random circuit suffices. This observation leads to an open problem: Are there symmetric Hamiltonians whose training samples could be generated by a shallow random circuit? If so, can they be trained on fewer samples?

Lastly, since quantum computation can be implemented on various platforms with different gate sets and connectivities, experimentalists must tailor their circuits to specific hardware constraints and capabilities. 
For a device with nearest-neighbor coupling, dynamical simulation of long-range interactions using Trotterization often requires additional compilation steps, leading to significant overhead due to the extensive use of SWAP gates. In contrast, PQCs offer greater flexibility by allowing the circuit to be trained with a desired architecture and gate set. In this work, we only considered $SU(4)$ gates with 1D and 2D brickwall connectivities, but it is of great interest to explore PQCs designed specifically for various platforms. For example, could certain quasi-local quantum dynamics benefit significantly from all-to-all connectivity, as in trapped ion quantum systems~\cite{malinowski2023wire}? Optimizing PQCs for such tailored connectivities could further enhance performance on specific hardware architectures.

\section{Acknowledgements}
We would like to thank Matteo D'Anna, Hong-Ye Hu, Tim Hsieh, Roger Luo, Andrew Potter, and Yifan Zhang for their insightful discussions.
RW would like to thank Max Hunter Gordon for his help in 2022 during the Quantum Computing Summer School in Los Alamos. J.C. acknowledges support from the Shared Hierarchical Academic Research Computing Network (SHARCNET), Compute Canada, and the Canadian Institute for Advanced Research (CIFAR) AI chair program. Y.Z., J.C., and Y.B.K. were supported by the Natural Science and Engineering Research Council (NSERC) of Canada. Y.Z. and Y.B.K. acknowledge support from the Center for Quantum Materials at the University of Toronto. Y.Z. was further supported by a CQIQC fellowship at the University of Toronto, and in part by grant NSF PHY-2309135 to the Kavli Institute for Theoretical Physics (KITP). LC was supported by Laboratory Directed Research and Development program of Los Alamos National Laboratory under project number 20230049DR. Resources used in preparing this research were provided, in part, by the Province of Ontario, the Government of Canada through CIFAR, and companies sponsoring the Vector Institute 
\url{https://vectorinstitute.ai/#partners}. The simulation code in this work can be
found at \cite{our_code}.

\bibliography{library.bib}
\appendix
\appendix
\onecolumngrid
\section{Learning dynamics via out-of-distribution generalization}\label{appx:learn}
In this section, we review the key quantum machine-learning theorems used in the main text. We start with the Hilbert-Schmidt norm,
\begin{align}
    \left\langle A, B\right\rangle_{HS} : = \frac{1}{N}\Tr[A^\dag B]
\end{align}
for $A,B\in\mathbb{C}^{N\times N}$. We see that the norm decomposes as follows:
\begin{align}
    \norm{A-B}_{HS}^2 &= \norm{A}^2_{HS} + \norm{B}^2_{HS} - \left\langle A, B\right\rangle_{HS} - \left\langle B, A\right\rangle_{HS}\nonumber\\
    &= \norm{A}^2_{HS} + \norm{B}^2_{HS} - 2\Re\left[\left\langle A,B \right\rangle_{HS}\right]
\end{align}
Note that for a unitary $U$ we have $\left\langle U, U\right\rangle_{HS}=\norm{U}^2_{HS} = N$. Hence for two unitary matrices $U$ and $V$, we find
\begin{align}
    \frac{1}{2N}\norm{U-V}_{HS}^2 = 1 - \frac{1}{N}\Re\left[\left\langle U,V\right\rangle_{HS}\right] 
\end{align}
Note that if we have $V=e^{i\phi} U$, we get
\begin{align}
    \frac{1}{2N}\norm{U-V}_{HS}^2 = 1 - \Re\left[e^{i\phi}\right]
\end{align}
If we choose the to ignore the global phase, a more natural cost function would be
\begin{align}
    C(U,V) = 1 - \frac{1}{N^2} \abs{\Tr[U^\dag V]}^2 .
\end{align}
We now take $V$ to be a parameterized unitary circuit $V(\btheta)$ with parameters  $\btheta = \{\theta_1,\ldots,\theta_d\}\in\mathbb{R}^d$.

Note that that the cost in Eq.~\ref{eq:un_fid} can be related to the average Haar fidelity 
\cite{horodecki1999teleport, nielsen2002simple}
\begin{align}
    C(U,V(\btheta)) &= \frac{d+1}{d}\left(1 - \Bar{F}(U,V(\btheta))\right)\\
     \Bar{F}(U,V(\btheta)) &=\mathbb{E}\left[\abs{\bra{\psi}U^\dag V(\btheta)\ket{\psi}}^2\right]_{\psi\sim \mathcal{P}_{\mathrm{Haar}}}\nonumber \\
     &= \int_{\psi} d\psi\: \abs{\bra{\psi}U^\dag V(\btheta)\ket{\psi}}^2
\end{align}
where $ \mathcal{P}_{\mathrm{Haar}}(\mathcal{H})$ is the uniform probability distribution of states in $\mathcal{H}$ and $d\psi $ is the Haar measure on $\mathcal{H}$.
Instead of using the Haar measure, we can write down a cost called the \emph{expected risk},
\begin{align}
    R_\mathcal{Q}(\btheta) = \mathbb{E}\left[1 - \abs{\bra{\psi}U^\dag V(\btheta)\ket{\psi}}^2\right]_{\psi\sim \mathcal{Q}} 
\end{align}
for some distribution of quantum states $\mathcal{Q}$. Note that $R_\mathcal{Q}$ has values $0\leq R_\mathcal{Q}\leq 1$ and that for $\mathcal{Q}= \mathcal{P}_{\mathrm{Haar}}$, we have $R_\mathcal{Q}(\btheta) = 1 - \Bar{F}(U,V(\btheta))$.
Given a data set $\mathcal{D}_\mathcal{Q} = \{\ket{\psi_k} \}$ of $K$ states $\ket{\psi_k} \sim \mathcal{Q}$, we can write down the \emph{empirical risk} function
\begin{align}
    C_{\mathcal{D}_\mathcal{Q}}(\btheta) = 1 - \frac{1}{K} \sum_{k=1}^K \abs{\bra{\psi_k}U^\dag V(\btheta)\ket{\psi_k}}^2 
\end{align}
which is an unbiased estimator for the true expected risk $R_\mathcal{Q}(\btheta)$. Note that this cost function is much easier to handle than Eq.~\ref{eq:un_fid} since we now have to measure the overlap between states sampled from $\mathcal{Q}$. In an ideal world, we would be able to sample from the Haar measure on $\mathcal{H}$, so that we get the equality $R_\mathcal{Q}(\btheta) = 1 - \Bar{F}(U,V(\btheta)) $, however, this is not possible in practice. The question is then, is there a ``simple" distribution $\mathcal{Q}$, s.t. $R_\mathcal{Q}(\btheta)$ approximates $R_{\mathcal{P}_{\mathrm{Haar}}}(\btheta)$? As show in \cite{caro2023learning}, the answer such distributions do indeed exist. First, we define the notion of a \emph{locally scrambling ensemble}.
\begin{definition} [Locally scrambling unitary ensemble] A unitary ensemble $\mathcal{U}_{LS}$ is locally scrambled iff for $U\sim\mathcal{U}_{LS}$ and for any fixed $U_1,\ldots, U_n \in \mathcal{U}(\mathbb{C}^2)$ also $U \cdot  (\bigotimes_{i=1}^n U_i) \sim \mathcal{U}_{LS}$. Here, $\mathcal{U}(\mathbb{C}^2)$ is the unitary Haar measure on a single qubit.
\end{definition}
In other words, acting with a random local unitary should to change the properties of the ensemble. We can use the above definition to define a locally scrambled state
\begin{definition} [Locally scrambling ensemble of states]
\begin{align}
    \mathcal{S}_{LS} = \mathcal{U}_{LS} \ket{0}^{\otimes n}.
\end{align}
\end{definition}
Given such a locally scrambling ensemble, the authors of \cite{caro2023learning} prove the following theorem:
\begin{theorem}[Equivalence of Locally Scrambled Risks]
    If $\mathcal{Q}$ is a locally scrambling ensemble, then
    \begin{align}
        \frac{1}{2} R_{\mathcal{P}_{\mathrm{Haar}}}(\btheta) \leq \frac{N}{N+1} R_{\mathcal{Q}}(\btheta) \leq R_{\mathcal{P}_{\mathrm{Haar}}}(\btheta).
    \end{align}
\end{theorem}
What this means is that any ensemble $\mathcal{Q}$ that is locally scrambling will give a risk that is within a factor $1/2$ of the empirical risk over the full Haar measure. 

The question we are then left with is: if we achieve a low empirical cost $C_{\mathcal{D}_\mathcal{Q}}$ on the data, how close to the actual expected risk $R_{\mathcal{P}}$ can we get? We have the answer with the following corollary from \cite{caro2023learning}.
\begin{corollary} \label{cor:ood_generalization}(Locally scrambled out-of-distribution generalization from in-distribution generalization)\newline
Let $\mathcal{P} \in \mathcal{S}_{LS}$ and $\mathcal{Q} \in \mathcal{S}_{LS}$. Let $U$ be an unknown $n$-qubit unitary. Let $V(\btheta)$ be an $n$-qubit quantum circuit with $T$ parameterized local gates. When trained using training data $\mathcal{D}_{\mathcal{Q}}$ containing $K$ input-output pairs, with inputs drawn from the ensemble $\mathcal{Q}$. Then, the out-of-distribution risk w.r.t. $\mathcal{P}$ with optimal parameters $\btheta^*$ after training satisfies
\begin{align}
    R_\mathcal{P}(\btheta^*) \leq 2C_{\mathcal{D}_{\mathcal{Q}}}(\btheta^*) + O\left(\sqrt{\frac{T\log T}{K}}\right).
\end{align}
\end{corollary}
Hence the true expected risk is of a locally scrambled ensemble can be upper bounded by the empirical cost and a factor that scales with the number of gates in the circuit and the number of states in the data.

\section{Training for systems with \texorpdfstring{$\mathrm{U}(1)$}{} symmetries}\label{app:u1}
\begin{figure}[htb!]
    \centering
    \includegraphics[width=0.9\linewidth]{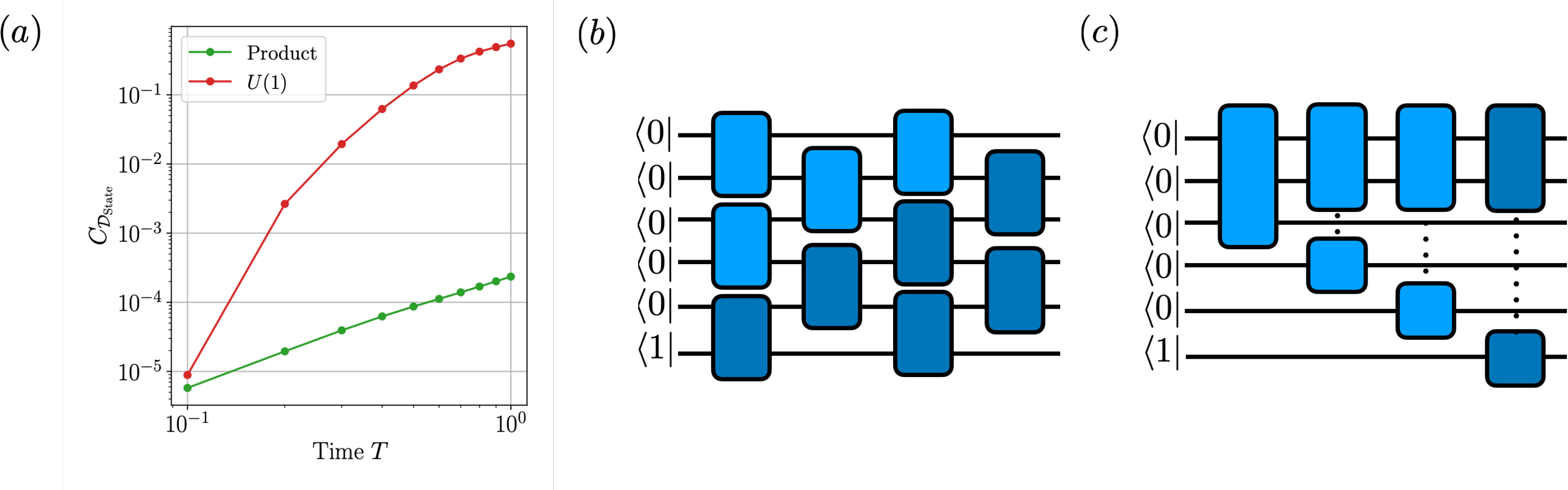}
    \caption{\textbf{Difficulty of generalization within $\mathrm{U}(1)$-preserving ensembles.} 
    (a) We compile a unitary of time $t=0.1$ and perform a long time evolution up to $T=1$ for the $n=20$ Heisenberg model. Note that this Hamiltonian has a $\mathrm{U}(1)$ symmetry to preserve the particle sector of a state under unitary evolution. As training data we use product states from $S_{\mathrm{Haar}^{\otimes n}}$ (green line) and $S_{\mathrm{U}(1)}$ within the half-filling charge sector(red line). We perform the evolution of a random $\mathrm{U}(1)$ state with half filling and calculate the overlap $C_{\mathcal{D}_{\mathrm{State}}}$ with TEBD simulation. We see that the generalization fails if we use training data from $S_{\mathrm{U}(1)}$. Only when the unitary is compiled with random product states as data do we observe the correct long-time dynamics. (b) We consider a $1d$ brickwall-shaped random quantum circuit (RQC) where each gate can be a $\mathrm{U}(1)$ charge conserving gate. In a shallow circuit, due to the light cone structure, the charge only remains inside the navy blue region, and thus the output state ensemble cannot be a locally scrambled one. (c) An MPS can be written as a sequential circuit~\cite{schon2005sequential,kim2017holographic,liu2019variational,soejima2020isometric,foss2021holographic,anand2023holographic,zhang2022qubit}, and a similar light-cone argument can be made.}
    \label{fig:u1_heis}
\end{figure}

A natural extension of the training protocol is considering a system with specific symmetries and restricting the training protocol to a specific symmetry sector. At first glance, training should be easier because the symmetry sector contains less information than the whole unitary. The next natural question is then: what training data set should we use? 

In this section, we consider the case of a system with a global $\mathrm{U}(1)$ symmetry.
This symmetry can be implemented with RQC or MPS. For computational efficiency, we require the data set being generated from an RQC with a fixed depth $\mathcal{D}_{\mathcal{Q}} \sim \mathcal{S}_{\mathrm{U}(1), \text{RQC}}$
or from a random MPS with a fixed bond dimension $\chi$ \cite{singh2011} $\mathcal{D}_{\mathcal{Q}} \sim \mathcal{S}_{\mathrm{U}(1), \text{MPS}}$. Here $\mathcal{S}_{\mathrm{U}(1)} = \mathcal{U}_{\mathrm{U}(1)} \ket{s}$ where $\ket{s}$ is a computational basis whose charge (Hamming weight) we want to preserve. In Fig.~\ref{app:u1}, we report the numerical results tested with the Heisenberg model. To our surprise, the generalization is actually \emph{worse} than the PQC trained on product state, even when tested on trial states within the same charge sector of the training data, see Fig.~\ref{fig:u1_heis}.

We can understand this result analytically. In fact, the failure of generalization is due to neither of these unitary ensembles being locally scrambling. 
A key requirement for the out-of-distribution generalization to work is that an ensemble shares the same expectation with the Haar random unitary ensemble when acted for all density matrices $\rho$, up to the second moment. 
\begin{align}
    \mathbb{E}_{U\sim \mathcal{U}_{\text {RQC}}}[U\rho U^\dag] = \mathbb{E}_{U\sim \mathcal{U}_{\text {Haar}_n}}[U\rho U^\dag]; \ \ \     \mathbb{E}_{U\sim \mathcal{U}_{\text {RQC}}}[U^{\otimes 2}\rho (U^\dag)^{\otimes2}] = \mathbb{E}_{U\sim \mathcal{U}_{\text {Haar}_n}}[U^{\otimes2}\rho (U^\dag)^{\otimes2}]
\end{align}
This however cannot be true in a shallow $\mathrm{U}(1)$ charge-conserving circuit. We sketch the proof for the RQC case but a similar argument can be made to the fixed-$\chi$ MPS case. 
\begin{theorem}[Shallow $\mathrm{U}(1)$ RQCs are not locally scrambling ensemble]
An RQC cannot form a locally scrambling ensemble with depth $\tau<n$.
\end{theorem}

Here we prove by contradiction:
\begin{proof}
    Consider $\rho$ is a product state $\ketbra{1000...}{1000...}$. The Haar random circuit would scramble the charge and thus the charge should spread evenly in the system; however, for the RQC, the charge spreading is limited by the circuit depth and the light cone of the $\ket{1}$ site. In order for the light cone to cover the whole system, the circuit depth has to be at least $n$.\\
    
    Therefore there exists a $\rho$ s.t. $\mathbb{E}_{U\sim\mathcal{U}_{\text {RQC}}}[U\rho U^\dag] \neq \mathbb{E}_{U\sim\mathcal{U}_{\text {Haar}}}[U\rho U^\dag]$ for any RQC depth $\tau<n$
\end{proof}
The training set generated by a shallow circuit therefore can not generalize well, matching what we observed numerically. Hence to achieve generalization within a $\mathrm{U}(1)$ charge sector, the training set has to be generated by a random circuit with depth at least $\tau = \Omega(n)$. Therefore, in all Heisenberg model training results, we did not use a U(1)-conserving state ensemble. 

\section{Optimization methods for PQC}
\subsection{Mitigatable barren plateaus in sample efficient learning}\label{app: bp}
Despite the success of variational compilation for many problems, the optimization can suffer from barren plateau (BP) issues. Roughly speaking, BPs occur when the landscape of the cost function has an average gradient variance that is exponentially small in the system size \cite{mcclean2018barren, ragone2023}. For local cost functions such as a single Pauli observable, this is known to happen in a $poly(n)$ circuit depth; whereas for global cost functions even $O(\log(n))$-depth circuits can be problematic to optimize \cite{Cerezo2021costfunctiondep}, which could prevent the succesful optimization of fidelity functions. 

One approach to mitigate BPs, similar previous works such as  ~\cite{khatri2019quantum} and \cite{jones2022robust}, is to consider an alternative local cost function:
\begin{align}
    C^{\rm local}_{\mathcal{D}_\mathcal{Q}}(\btheta) = 1 - \frac{1}{K} \sum_{k=1}^K\frac{1}{n}\sum_i^n\text{Tr}[(\ketbra{\psi_{k,i}}\otimes\mathbb{I})V^\dag(\btheta)U\ketbra{\psi_k}U^\dag V(\btheta)],
    \label{eq:local_risk}
\end{align}
where $\ketbra{\psi_{k,i}}$ is the reduced density matrix of $\ketbra{\psi_{k}}$ on site $i$. When Eq.~\ref{eq:local_risk} equals $0$, Eq.~\ref{eq:erisk} is guaranteed to be $0$. On paper, it seems like Eq.~\ref{eq:local_risk} reduces the global cost function to a local cost function where BPs are less severe. However, in practice, this introduces an extra summation over each site $i$ in the cost function, which adds extra computational costs. Some work reduces the contraction cost by restricting the cost function to the lightcone of the targeted evolution~\cite{mizuta2022local}. However, we found that this cost is still too prohibitive for our numerical studies. Furthermore, it is known that the learning is less efficient if it is not ``end-to-end"~\cite{luo2024operator},  meaning that when the target cost function is not directly aligned with the desired outcome, the learning process becomes less effective. 

\begin{figure}[htb!]
    \centering
    \includegraphics[width=0.45\linewidth]{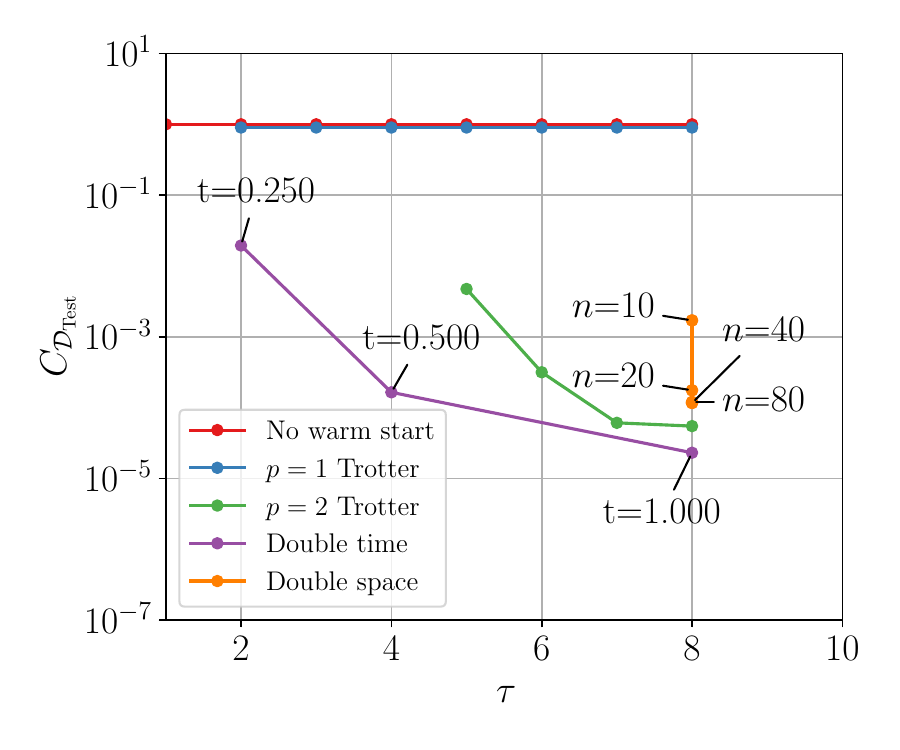}
    \caption{\textbf{The importance of warm starts.} We show the results of a variational compilation procedure for the one-dimensional Ising model at $g=1$ for $n=80$ and $t=1.0$. All circuits are translation invariant (same parameters for all gates per layer) and trained with $N_s = 16$.} 
    \label{fig:init}
\end{figure}

Nevertheless, the time evolution we are trying to learn here has a lot of structure that can be exploited. A series of ``warm start" strategies have been proposed and proven successful in synthesizing quantum dynamics. For example, one could truncate the time evolution into slices. Starting from a single time slice and iteratively ``grow" the dynamics, using the parameters from a previous iteration as the initialization~\cite{lin2021real}. In a recent work~\cite{drudis2024variational}, the authors analytically study this iterative initialization strategy and prove such a variational algorithm has at least an inverse polynomial gradient in $n$ at each time step. By establishing convexity guarantees for these regions, their work suggests polynomial-size time steps could be trainable. While the initialization strategy considered in~\cite{drudis2024variational} is successful, it requires iterations linear in the total evolution time and can be costly to implement numerically.
In this work, we use three different initialization strategies to overcome BPs, inspired by ideas in transfer learning~\cite{zhuang2020comprehensive}:
\begin{enumerate}
    \item \emph{Trotterization.}
    We start the optimization from an order $p$ Trotterized time evolution. This has the advantage that it is completely deterministic and gives us control over the starting fidelity of the optimization. This has the benefit that it is completely deterministic. However, sometimes Trotterization does not give a circuit that matches the desired PQC architecture so other initialization methods are considered in our work.
    \item \emph{Double time.}
    Here, we start with a very short time evolution $t_{\mathrm{init}}$ with a circuit of depth $\tau_{\mathrm{init}}$. For short times, even for large systems, this should result in a finite overlap between the training states $U\ket{\psi_i}$ and targets $V_{\mathrm{init}}(\btheta)\ket{\psi_i}$. After optimization, we obtain the parameters $\btheta^*$. Then, we double the time $t_1 = 2t_{\mathrm{init}}$, and initialize the circuit with two times the previous solution: $V_1(\btheta) = V_{\mathrm{init}}(\btheta^*)V_{\mathrm{init}}(\btheta^*)$. This initialization will again have a non-zero fidelity, hence we can re-optimize $\btheta$. We repeat this process until we reach some target time $t_{\mathrm{final}} = t_{\mathrm{init}} \times 2^i$.
    \item \emph{Double space.}
    Similar to the ``Double time" strategy, we start with a system size $n_{\mathrm{init}}$ for a fixed time $t$ and fixed depth $\tau$. We solve the optimization problem and find a solution $V_{\mathrm{init}}(\btheta^*)$. Then, we double the system size, $n_1 = 2 n_{\mathrm{init}}$ and initialize the new circuit with the tensor product of the previous solution: $V_1(\btheta) = V_{\mathrm{init}}(\btheta^*)\otimes V_{\mathrm{init}}(\btheta^*)$. We re-optimize the parameters $\btheta$ for this new system size and keep repeating this procedure until we reach a target system size $n_{\mathrm{final}} = n_{\mathrm{init}} \times 2^i$.
\end{enumerate}
In Figure \ref{fig:init}, we show how these three different initialization strategies behave in a typical setting. First, we see that without any warm start, the optimization will get stuck at $C_{\mathcal{D}_{\mathrm{Test}}}=1$. Similarly, a $p=1$ trotterization is not enough to get the optimization going. Only when we initialize a $\tau=5$ circuit with a $p=2$ Trotterization, do we get a large enough signal to start the optimization. We see that both the ``Double time" and ``Double space" approaches result in large enough signals to start the optimization. Interestingly, we see that different warm start approaches lead to different solutions that can vary almost an order of magnitude in infidelity.
\clearpage
\subsection{Global update versus local update}\label{appx:update}

In our study, we implemented a global parameter update scheme where all variational parameters in the PQCs were updated simultaneously. Here we compare this approach to a local update scheme, which is akin to the density matrix renormalization group (DMRG) method~\cite{white1992density}, and has been considered in recent works such as those by Lin et al. and Drudis et al.~\cite{lin2021real,drudis2024variational}. In the local update method, the PQC is sequentially optimized by adjusting one layer of tensors at a time, sweeping from left to right and then back again, and this process is repeated until convergence is achieved.

Our results indicate that for shallow circuits (depth $\tau \leq 4$), the outcomes of the local update scheme are comparable to those of the global update scheme. However, as the circuit depth increases, the local update method struggles to locate the global minimum, particularly at a depth of $\tau = 8$, where the global update outperforms the local update by more than an order of magnitude. This highlights a key difference between ground state finding problems and our scenario, suggesting that local updates may not always be effective in achieving the global optimum in more complex optimization landscapes.

\begin{figure}[htb!]
    \centering
    \includegraphics[width=0.75\linewidth]{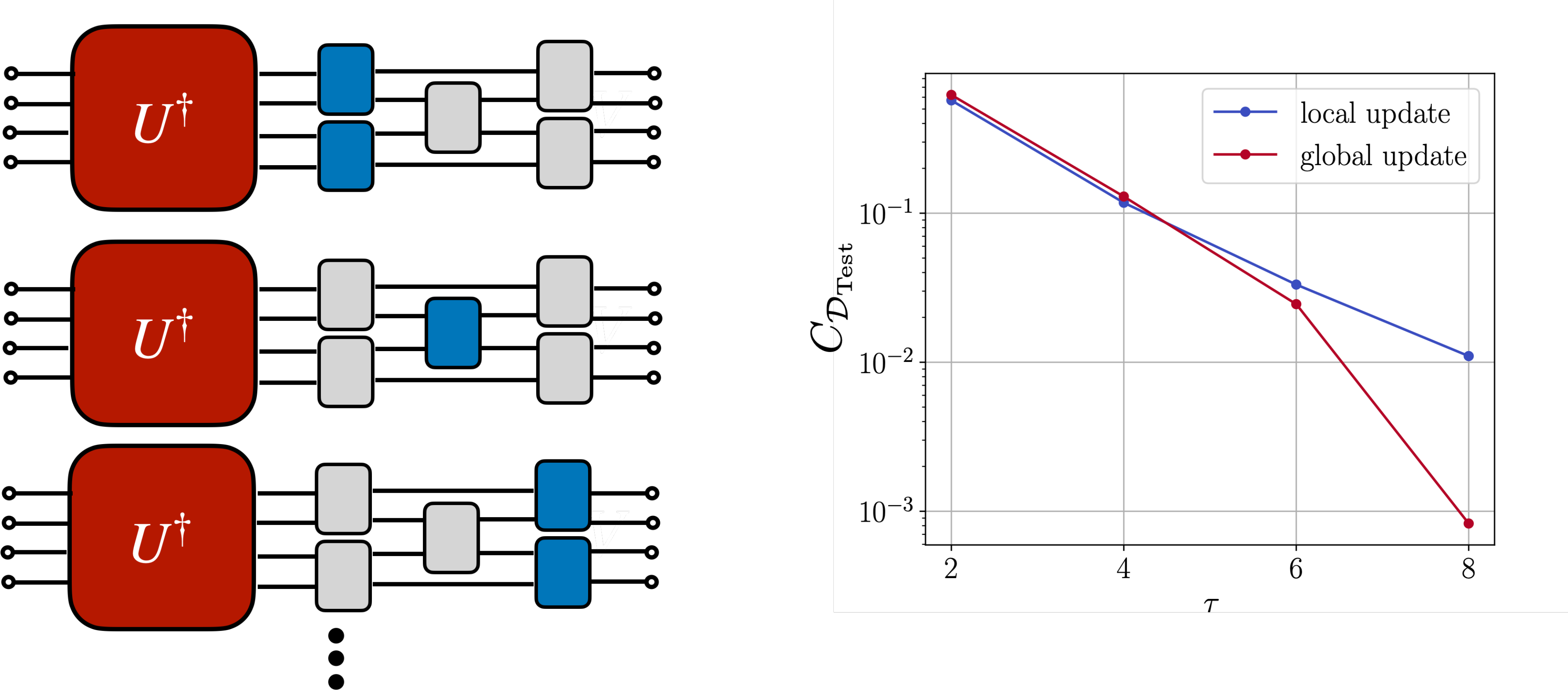}
    \caption{\textbf{Comparing local and global update schemes} Left: A demonstration of the local sweeping scheme. In this method, the PQC is translation invariant, and only one layer of the circuit is updated at a time. Right: Performance comparison of the two update schemes on a $4\times4$ Ising model with parameters $t=0.5$, $g=-1$, and $\kappa=0$.}
    \label{fig:update}
\end{figure}
\begin{table}[]    
\centering
    \begin{tabular}{c|c}
        Gate & \#CNOTs \\\hline
        SWAP & 3\\
        nearest-neighbor $XX$-Gate & 2\\
        next nearest-neighbor $XX$-Gate & 6\\
        $\mathrm{SU}(4)$ & 3\\ 
    \end{tabular}
    \caption{A list of the (nearest neighbor) CNOT costs of gates considered in this work. For the Ising Hamiltonian, the $XX$ gate costs only $2$ CNOTs instead of $3$. The non-local gate requires us to use a SWAP, which triples the CNOT count.}
    \label{tab:cnot_costs}
\end{table}

\end{document}